\documentclass{eas}
\usepackage{graphicx}
\usepackage{sidecap}
\begin{document}

\title{Tidal Flows in asynchronous binaries: The $\beta$-factor} 
\runningtitle{The $\beta$-factor}
\author{Gloria Koenigsberger and Edmundo Moreno}\address{Instituto de Astronom\'{\i}a,
                 Universidad Nacional Aut\'onoma de M\'exico,
                 Apdo. Postal 70-264
                 D.F., 04510  M\'exico}
%
%
\begin{abstract}

We discuss the potential role that tidal flows in asynchronous binary stars may
play in transporting  chemically enriched material from deep layers towards the surface 
and the corresponding observational consequences of these processes. We suggest
that the time-dependent velocity field induced by asynchronous rotation may contribute 
significantly to the mixing, thus providing a channel for the formation of chemically
enriched slowly rotating massive stars. 
\end{abstract}
\maketitle

\section{The $\beta$-factor}

Models of binary stars are often based on the assumption that the star is in an equilibrium state; i.e., 
that it is in a circular orbit ($e$=0),  rotates at a rate $\omega$ equal to the rate
of orbital revolution $\Omega$ and that the equator coincides with the orbital plane (Hut 1981).
In such an idealized state, the deformation induced by the tidal forces remains constant throughout the 
orbital cycle and there are no relative motions between the stellar layers. Hence, 
$\omega$=$\omega (r', \varphi', \theta',t)$= $\omega_0$=constant for each point (r', $\varphi'$, $\theta'$)
in the star and at all time $t$ throughout the orbital cycle.  If any one of the above conditions is 
relaxed, a time-dependent velocity field is generated and $\omega$ no longer is constant. Defining
$\beta$=$\omega/\Omega$, the equilibrium state is described with $\beta$=1.

A qualitative picture for the non-equilibrium cases ($\beta \neq$1) can be drawn 
by adopting a simplified model in which the star is divided into an outer layer that is free to move 
in response to the forces that are applied to it and an inner rigid body rotating with angular 
velocity $\omega_0$ upon which this layer rests (Fig. 1).  In the general case
of an eccentric binary system, one can then define $\beta_0$=$\omega_0/\Omega_0$, where $\Omega_0$ is 
the magnitude of the orbital angular velocity vector at some reference point, which we adopt as 
the time of periastron. 



\section{Tidal flows}

Tidal flows refer to the velocity field that appears when $\beta \neq$1.
Here we shall focus on the azimuthal component of the tidal velocity field, v$_{\varphi '}$,
since its contribution has the potentially most significant impact in creating conditions by which 
different stellar layers interact and may thus impact the mixing processes.
Analytical expressions for v$_{\varphi '}$ have been derived only under simplifying approximations,
for example, assuming that  $\beta \simeq$1, and viscosity $\nu \rightarrow$0 (Scharlemann 1981). 
The  complexity of the general problem resides in the number and magnitude of the forces that
are involved:  In addition to gravity, one must include  the centrifugal and Coriolis effects, 
the hydrodynamic properties of the stellar material and viscous shear; and the full solution 
requires a 3D calculation. 

Considerable insight can be gained, however, through a method based on the solution of the equations of 
motion for an $n \times m$ grid of surface elements as shown in Fig.~1.  In exchange for the 
surface-layer simplifying approximation, it is possible to include all the forces mentioned above and 
perform the calculation for arbitrary rotation velocity,  orbital eccentricity and viscosity.  This method
is implemented in the TIDES code (Moreno \& Koenigsberger 1999; Moreno et al. 2011) making it possible 
to examine the behavior of v$_{\varphi'}$ for different regimes, such as those of  sub-synchronous or 
super-synchronous rotation and highly eccentric orbits.  The most important result of the many 
calculations we have performed is that when $\beta\neq$1, v$_{\varphi'}$ is always non-uniform
over $\varphi'$.  This is in stark contrast with the differential rotation profiles generally
adopted for single stars (i.e., the ``shellular'' approximation; see, for example, Meynet \& 
Maeder 1997). On the other hand, the presence of accelerations and decelerations in the azimuthal 
direction is known from hydrodynamical calculations of rotating fluids in the presence of an external
gravitational field (Tassoul 1987).

\begin{SCfigure}  
\includegraphics[width=0.30\linewidth]{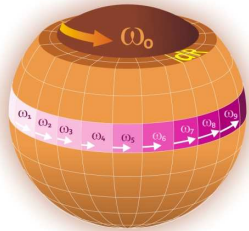}
\caption{The TIDES one-layer model  assumes an interior rigid body rotating at a 
rate $\omega_0$ and an outer layer of thickness $dR$ that responds to gravitational, 
centrifugal, Coriolis, gas pressure and viscous forces.  
}
\end{SCfigure} 

As an illustration, consider a binary star of mass $m_1$=5 M$_\odot$ and equilibrium radius 
$R_1$=3.2 R$_\odot$ which has a companion of mass $m_2$=4 M$_\odot$.  As in Moreno \& Koenigsberger (1999), 
we  adopt a rotating reference frame centered on $m_1$ that revolves at a rate $\Omega$, which
coincides with $m_2$'s orbital angular velocity.  The azimuthal coordinate $\varphi '$ is measured from 
the line that connects $m_1$ and $m_2$,  and $\theta '$ is the co-latitude angle.  Fig.~2 shows the 
behavior of v$_{\varphi'}$  obtained with our TIDES code (Moreno et al. 2011) for two binary systems:  
{\em Model-1} was computed with $P$=2d and eccentricity $e$=0 , and  {\em Model-2} with $P$=14d and $e$=0.7.  
In both cases, $\beta_0$=1.2 and $\nu$=0.003 R$^2_\odot$/day.

For {\em Model-1},  v$_\varphi'$  has two maxima and two minimal for all orbital phases.  
For {\em Model-2}, the number and amplitude of maxima change over orbital phase.  In both cases, 
the variability amplitude declines from the equator to the poles. Although TIDES does not compute 
the behavior of v$_{\varphi'}$ at different depths, it is clear that its amplitude must decrease in 
deeper layers (see, for example, Dolginov \& Smel'chakova 1992).  Thus, when $\beta\neq$1, the effect 
of tidal flows on the differential rotation structure is highly non-uniform and time-dependent. 

\begin{figure} [!h]
\includegraphics[width=4cm]{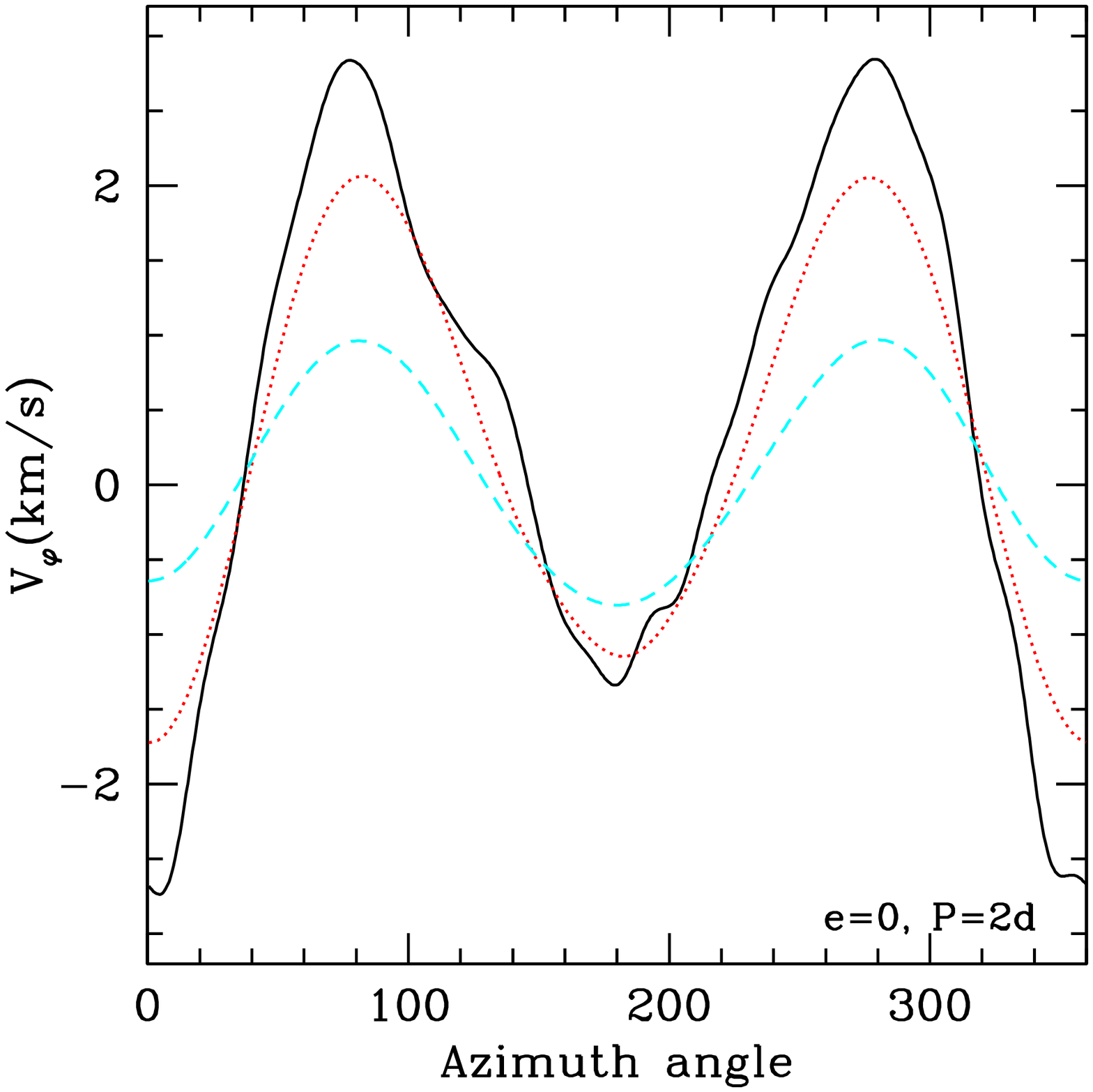}
\includegraphics[width=4cm]{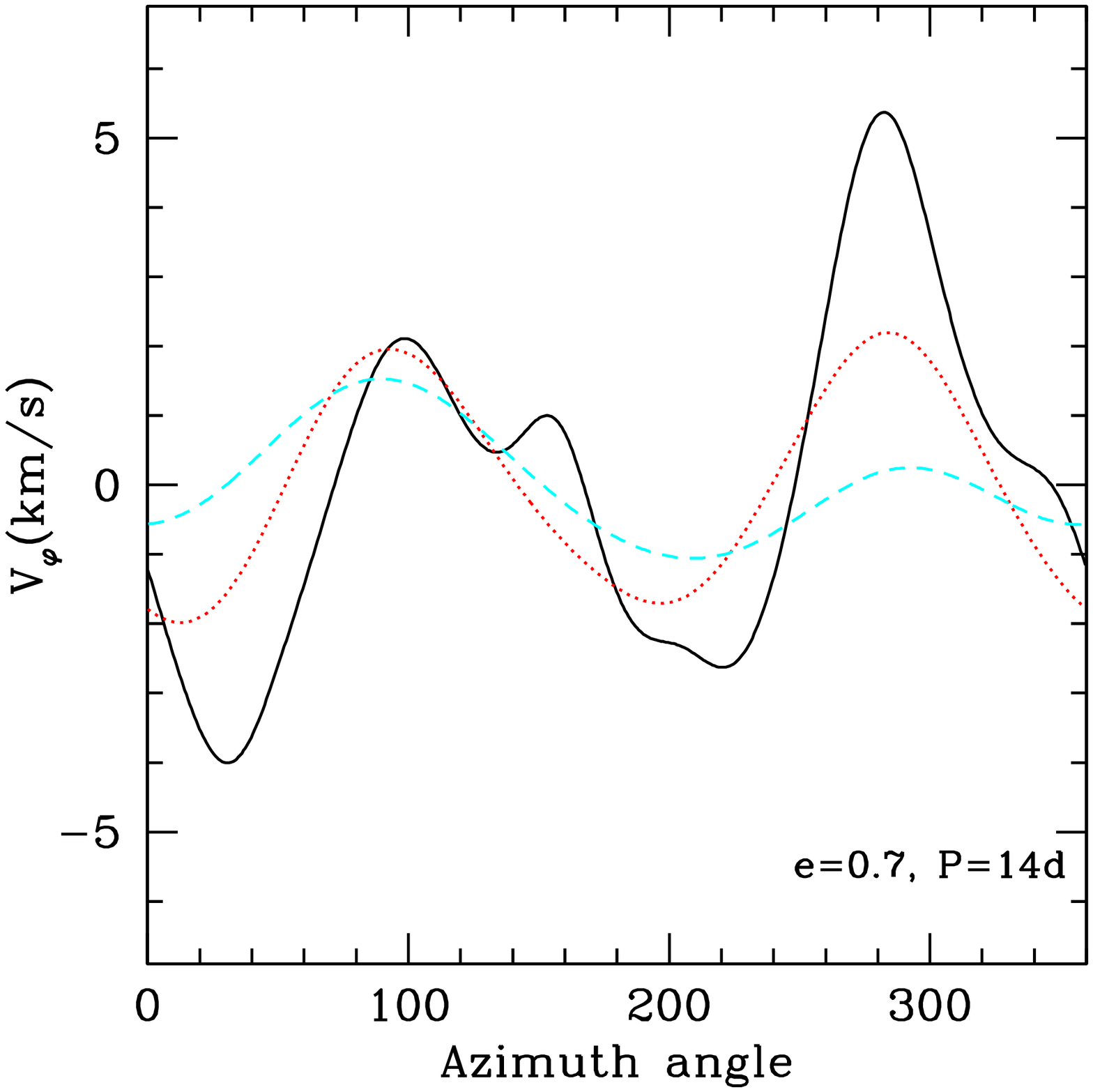}
\includegraphics[width=4cm]{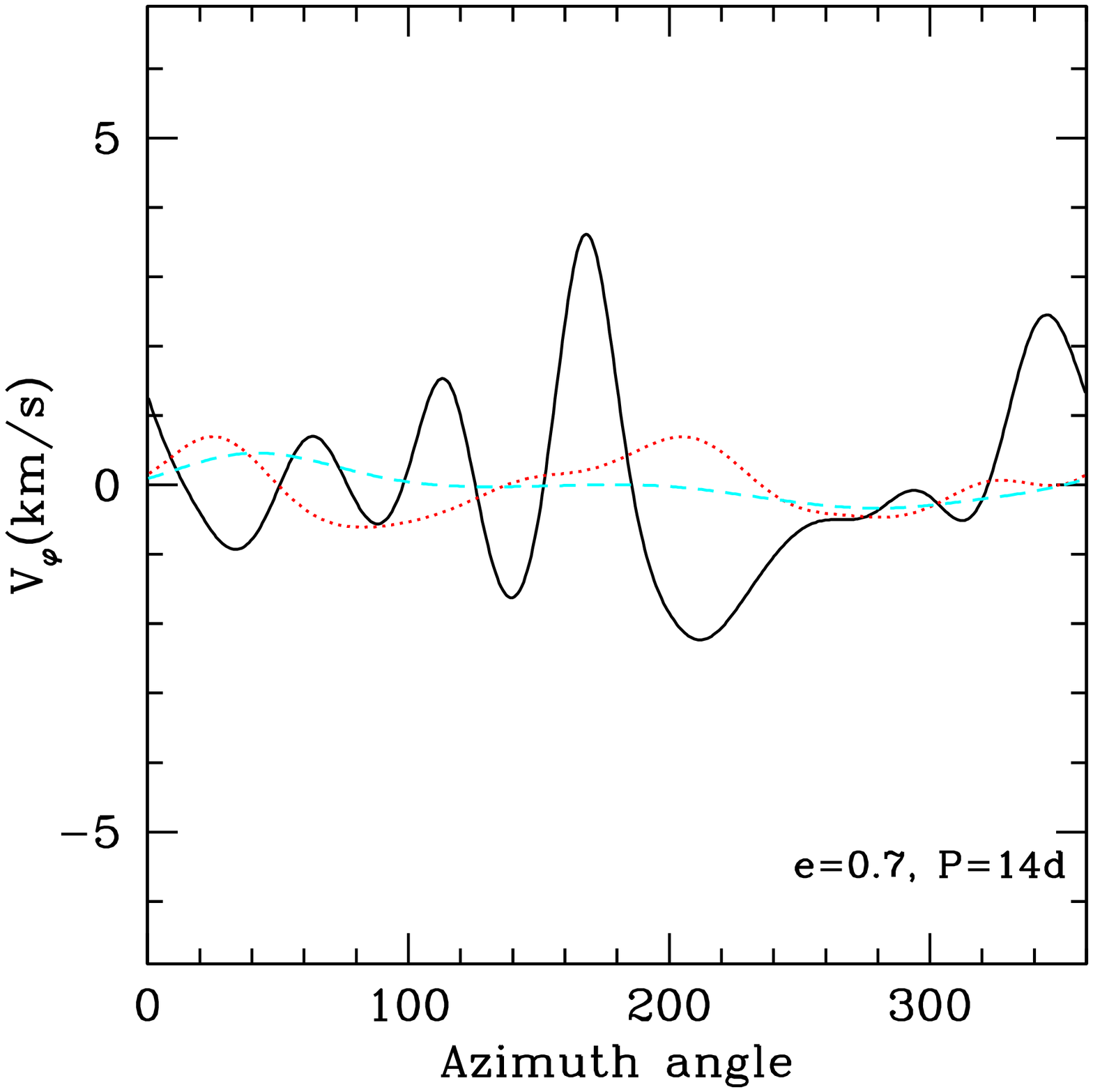}
\caption{Surface azimuthal velocity field at co-latitudes 90$^\circ$ (equator; continuous),
64$^\circ$ (dot) and 46$^\circ$ (dash) plotted against azimuth angle in the reference frame
rotating with the companion. Here $\varphi'$=0 at the sub-binary longitude and negative values
correspond to perturbations in the direction opposite to that of rotation.  {\bf Left:} {\em Model-1}; 
{\bf Middle:} {\em Model-2} at periastron; {\bf Right:} {\em Model-2} at apastron.
}
\end{figure}

\section{Potential consequences}

Stellar evolution models predict a correlation between rotation and mixing (Meynet \& Maeder 2000;
Brott et al. 2011), since the coupling of differentially rotating layers contributes to transport 
nuclear-processed elements from the stellar core to the surface.  Hence, rapidly rotating
stars are expected to have larger heavy-element enrichment on their surfaces than
stars rotating more slowly.  However, recent observational results have shown some discrepancies with 
this prediction since a significant number of slow rotators have been found to have large
Nitrogen surface abundance enrichment while, at the same time, some very fast rotators lack the expected enrichment
(Hunter et al. 2008).

Including the effects due to tidal flows may explain the discrepancies if a large number of
the outliers in the {\em N-abundance} {\it vs.} {\em rotation velocity}  diagram are binaries. 
Consider a binary star with negligible 
rotation ($\beta \sim$0) throughout most of its main sequence lifetime. The v$_\varphi'$ distribution 
in this case is highly non-uniform, as illustrated in Fig.~3, and assuming that tidal flows contribute  
significantly towards the  mixing, this star would be a non-rotating object with enriched heavy element abundances
on its surface.  Conversely, consider a very short-period system that attained equilibrium ($\beta_0$=1) 
very early during its main sequence lifetime.  From a strict theoretical standpoint, this condition 
implies that no significant differential rotation should be present and hence, the mixing that 
could be induced from this process does not take place.  Thus, this star would be a rapid rotator 
(v$_{rot}\sim$100 km s$^{-1}$) with little or no enrichement of heavy element abundances on its surface. 

Another interesting consequence involves the evolutionary paths that may be followed by
the two members of a binary system in the case that each star has a different $\beta$-value.  
For example, if $\beta(m_1)$=1 while $\beta(m_2)\neq$1, the above considerations suggest that
$m_2$ would evolve with significantly more mixing than $m_1$ and thus follow a different
evolutionary path.  

Many questions need to be addressed to assess the relative importance of
tidal flows with respect to the other mixing processes.
Key issues concern the value that is adopted for the viscosity parameter,
since it provides the coupling between the differentially-moving stellar layers; and the fact
that the interaction between tidal flows and, for example, convection is a 3D problem in which
very different timescales prevail.

\vskip0.2cm
\noindent {\it Acknowledgements}: We thank Raphael Hirschi and Fr\'ed\'eric Masset for helpful discussions and Francisco Ruiz
for computing assistance.  This work is supported by UNAM/PAPIIT grant 105313 and CONACYT grant IN129343.

\begin{SCfigure}  
\includegraphics[width=0.48\linewidth]{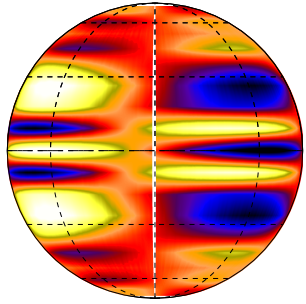}
\caption{The v$_{\varphi'}(\varphi',\theta')$ distribution over the surface of a $m_1$=61 M$_\odot$,
$R_1$=10.6 R$_\odot$ non-rotating star ($\beta_0$=0) with a $m_2$=66 M$_\odot$ companion in
a P=19.3d circular orbit and  viscosity $\nu$=4.5$\times$10$^7$ cm$^2$ s$^{-1}$.  Light (dark) colors 
represent flows in the same direction  (opposite direction) to that of stellar rotation. The white line 
indicates the sub-binary longitude.
}
\end{SCfigure}




\begin{thebibliography}{99}
\bibitem[1992]{dolginov2} Dolginov, A.Z. \& Smel'chakova, E.V. 1992, A\&A, 257, 783.
\bibitem[2008]{hunter}   Hunter, I., Brott, I., Lennon, D.J. \etal\ 2008, ApJ, 676, L29.
\bibitem[1981]{hut} Hut, P. 1981, A\&A, 99, 126.
\bibitem[1997]{meynet} Meynet, G. \& Maeder, A. 1997, A\&A, 321, 465.
\bibitem[2000]{maeder}  Maeder, A. \& Meynet, G. 2000, ARA\&A, 38, 143.
\bibitem[1999]{moreno1} Moreno, E. \& Koenigsberger, G. 1999, RMAA, 35, 157.
\bibitem[2011]{moreno2} Moreno, E., \& Koenigsberger, G. \& Harrington, D.M. 2011, A\&A, 528, 48.
\bibitem[1981]{scharleman} Scharlemann, E.T. 1981, ApJ, 246, 292.
\bibitem[1989]{tassoul} Tassoul, J.L. 1987, ApJ, 322, 856.



\end{thebibliography}
\end{document}